# Effect of a weak longitudinal modulation in refractive index on transverse localization of light in 1D disordered waveguide lattices

Somnath Ghosh, R. K. Varsheny and Bishnu P. Pal
Physics Department, Indian Institute of Technology Delhi, Hauz Khas, New Delhi 110016, India
* somiit@rediffmail.com

*Abstract*—**We report the enhancement of the effect of transverse localization of light (TL) in presence of a weak longitudinal modulation of refractive index in disordered waveguide lattices. In our chosen lattices, tunneling inhibition along length favors to achieve the diffraction-free propagation along with the simultaneous presence of transverse disorder. Results will be useful to tune the threshold value of disorder to achieve localized light.**

## I. INTRODUCTION

After the prediction of the phenomenon of Anderson localization of wave functions in disordered systems in 1958 [1] in the context of solid-state physics which includes possibility of localization of light in certain disordered dielectric lattices, this effect had regained its momentum as one of the most evolving area in physics in 1989 [2]. Raedt et al. [2] studied the concept of Anderson localization of light in the presence of transverse disorder in a semi-infinite disordered dielectric geometry, and introduced the scheme of transverse localization (TL). However this scheme was first experimentally implemented in 2007 [3], which was the first experimental observation of Anderson localization in any periodic system containing disorder in its original context. Numerous works followed up this scheme [3-5] and in recent years, there has been a resurgence of interest in this area from fundamental physics point of view, as well as applications in photonics like disordered/random lasing in disordered medium [6,7]. Various platforms are being investigated in all three dimensions of multilayer photonic structures to realize these disordered lasing systems. Moreover, tuning the threshold value of the required disorder to achieve localized state in such disordered lattices is another key issue to look into to check the suitability of these optical systems for this particular application. To address this issue one needs to revisit the fundamental aspects of the localization effect. One key requirement to achieve localization is to maintain the refractive index pattern which includes the periodic backbone along with transverse disorder, unchanged along the direction of propagation. On the other hand, tunneling inhibition among different channels of any discrete photonic structures by longitudinally oscillating refractive index is often referred as dynamic localization (DL) [8]. Hence, synthesis of these two fields in the perturbation limit may lead to interesting propagation effects in disorder-affected periodic optical structures and open up a new possibility to tune the threshold for localization. In this paper, in our chosen specific 1D coupled waveguide lattice configuration, we have incorporated bi-directional modulation of refractive index, i.e. along both transverse and longitudinal directions. The depth and period of the longitudinal variation is chosen such a way that it acts only as a perturbation to the original effect of transverse localization of light. Our initial findings have established the fact that in a weak dynamically modulated lattice, a faster transition to the localized regime is possible even after a relatively shorter distance of propagation and also enhances the quality of localization.

## II. DISORDERED WAVEGUIDE LATTICES

To design a discrete photonic system to study light localization, we consider an evanescently coupled waveguide lattice consisting of a large number ($N$) of unit cells, and in which all the waveguides spaced equally apart are buried inside a medium of constant refractive index $n_0$ [4,5,9,10]. The overall structure is homogeneous in the longitudinal ($z$) direction along which the optical beam is assumed to propagate, as shown in Fig. 1(a). The change in refractive index $\Delta n$ (x) (over the uniform background of $n_0$) due to disorder in this 1D waveguide lattice is assumed to be of the form

$$\Delta n(x) = \Delta n_p (H(x) + C\delta(x)) \quad (1)$$

where $C$ is a dimensionless constant and it represents the level/strength of disorder. The periodic function $H(x)$ is 1 inside the higher index regions and is zero otherwise. $\Delta n(x)$ consists of a deterministic periodic part $\Delta n_p$ of spatial period $\Lambda$ and a random component $\delta$ (uniformly distributed over a specified range varying from 0 to 1) with the same spatial periodicity. With this particular choice of randomly varying refractive indices in the high index layers, which are also separated by randomly chosen layers (over and above an average spatial periodicity about the center), we have modeled the diagonal and off-diagonal disorders to study localization [4,5]. Wave propagation through the lattice is governed by standard scalar Helmholtz equation, which under paraxial approximation could be written as [3]

$$i\frac{\partial A}{\partial z} + \frac{1}{2k}\left(\frac{\partial^2 A}{\partial x^2}\right) + \frac{k}{n_0}\Delta n(x,z)A = 0 \quad (2)$$

where $A(x, z)$ is amplitude of an input CW optical beam having its electric-field as $E(x,z,t) = Re[A(x,z)e^{i(kz-\omega t)}]$ ;

$k = n_0 \omega / c$ .whose refractive index profile is assumed to be of the form
$$\Delta n(x, z) = \Delta n(x)[1 + \mu \sin^2(\Omega z)] \quad (3)$$

here μ represents the longitudinal modulation depth and Ω is the modulation frequency. We solve Eq. (2) along with Eqs. (1) and (3) numerically through the scalar beam propagation method, which we implemented in Matlab.

Our chosen waveguide lattice consists of 100 evanescently coupled waveguides as shown in Fig.1(a) (in absence of any modulation along length), each of 7 μm width and separated from each other by 7 μm. The value of $\Delta n_p$ was chosen to be 0.001 over and above the background material of refractive index $n_0 = 1.46$. In the case of a CW beam having a plane-phase front as the input Eqn. (2) yields localized exponential solutions [2]. Fig. 1(b) shows the modulation of the transverse index profile along the lattice length for a chosen spatial periodicity and modulation depth. We have shown the schematic representation of synthesis of two different light localization effects in the perturbation limit in Fig 1(c).

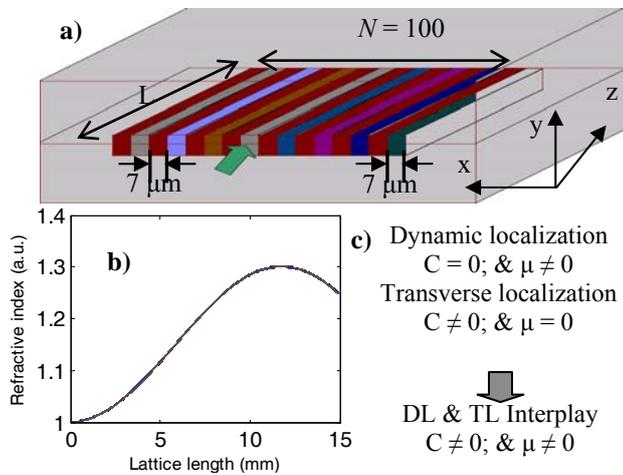

Fig. 1. (Color online) Schematic of a 1D coupled disordered lattice. Different shades of color signify different refractive indices over an average refractive index. b) Depth of longitudinal index modulation (normalized with respect to 1.46) along the lattice sample. c) Transverse localization in the presence of weak dynamic modulation which acts as a perturbation: bi-directional modulation of index profile.

### III. TRANSVERSE LOCALIZATION OF LIGHT IN MODULATED DISORDERED LATTICES

To achieve transverse localization of light in a disordered optical system, we study the propagation of an input CW beam with a plane phase front at an operating wavelength of 980 nm in our chosen disorder lattices. Various strengths of transverse disorder are controlled by choosing different *C* values as prescribed in Eq. (1). The input beam is assumed to cover few lattice sites around the central unit cell at the input plane; chosen beam width $\omega_0$ (FWHM) was 10 μm (> width (7 μm) of an individual lattice site).

The results in terms of averaged output beam width are shown in Fig. 2. A measure of the localization is assumed to be quantifiable through decrease in the average effective width ($\omega_{eff}$) (as defined in [5,11])
$$P \equiv \left[\int I(x,L)^2 dx\right] / \left[\int I(x,L) dx\right]^2$$
$$\omega_{eff} = \langle P \rangle^{-1} \quad (4)$$

of the propagating beam. As the propagation distance is always finite in practical systems, the measurement/estimation of $\omega_{eff}$ is an expectation value problem. Hence the statistical nature of the localization phenomenon has been incorporated through the ensemble averaging. It can be seen that in the case of a modulated lattice (μ = 0.3 & Ω = 0.002), the width of a localized state is much smaller than its counterpart in unmodulated (μ = 0) lattice when the strength of disorder is set at 0.4 in both the cases. Moreover, Fig. 2 clearly indicates that in a modulated disordered lattice a faster transition to the localized regime is possible even after a relatively shorter distance of propagation.

After a certain propagation distance, depending upon the strength of disorder and lattice aspect ratio, $\omega_{eff}$ becomes almost unchanged with characteristic statistical fluctuations. It is worthwhile to point out that physically as we increase the value of *C* parameter, the total number of localized eigen-states supported by a particular realization of the disordered lattice increases for a given length [4], which eventually favors a smaller $\omega_{eff}$ value. Naturally, a larger value of $\omega_{eff}$ in the localized regime as well as an increase in the localization length [3,11] of an eigen-state in a disordered medium would imply degradation in the quality of localization. This plot of $\omega_{eff}$ (ensemble averaged over 100 different realizations of a particular level of disorder and normalized with respect to $\omega_0$) with *L* in Fig. 2 clearly reveals enhancement in the quality of localization with propagation in the presence of a weak longitudinal modulation in the chosen disordered waveguide lattice.

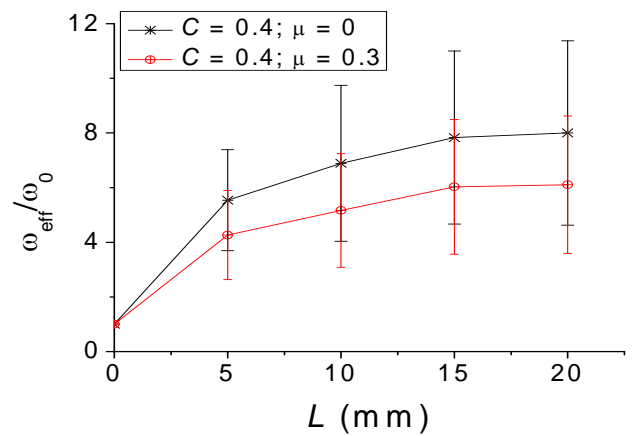

Fig. 2. (Color online) Variation in the ensemble averaged effective width of the output beam for an input Gaussian beam (FWHM 10 μm) for various propagation distances with two different longitudinal modulation depths at a disorder level of 40%. The error bars are the statistical standard deviations of beam widths. It shows a faster transition to localization in a modulated lattice.

To appreciate the above-mentioned enhancement effect, in Fig. 3 we have plotted the ensemble-averaged (over 100

realizations) $\omega_{eff}$ after a propagation through 15 mm long disordered waveguide lattice of the above kind for $\mu = 0$ as well as a finite $\mu$ set at 0.3 when the level of disorder ($C$) is varied from 0 to 0.6. The plots in Fig. 3 shows the transition from ballistic to localized regime of propagation along with an intermediate stage in which both diffraction and localization are simultaneously present, before it attains a localized state. If we compare the plots corresponding to finite $\mu$ relative to the plot for $\mu = 0$, it is obvious that a finite $\mu$ essentially favors localization to occur irrespective of the strength of disorder. However, if the disorder is sufficiently strong then the effect of TL takes over. Hence, it can be concluded that simultaneous presence of a weak longitudinal modulation along with transverse disorder in refractive index would favor a faster transition to localization.

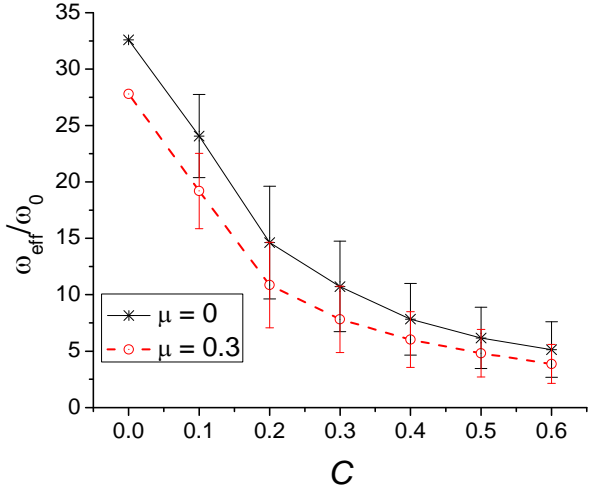

Fig. 3. (Color online) Variation in the ensemble averaged effective width of the output beam for an input Gaussian beam (FWHM 10 μm) with different strengths of disorder after a propagation distance of 15 mm. The error bars are the statistical standard deviations of beam widths. It confirms a faster transition to localization in a modulated lattice.

For a deeper appreciation of this effect on the degree of localization, in Fig. 4, we have plotted the ensemble averaged (over 100 localizations) output intensity profiles from 15 mm long disordered waveguide lattice. A particular localized state carries the signature of a corresponding disordered optical system. These plots also confirm the advantage of simultaneous presence of weak longitudinal modulation from localization point of view. We estimate the so-called localization length ($l_C$) characteristic of a localized state in a particular disordered lattice [3]. To obtain $l_C$, we averaged 100 output intensity profiles for a given value of $C$ and then performed a three-point moving average to smoothen further the resulting profile as mentioned in [11]. The estimated value of $l_C$ for the modulated ($\mu = 0.3$) disordered ($C = 0.5$) lattice is nearly 13% less relative to its counterpart unmodulated disordered lattice. It is also obvious that larger the absolute value of $\mu$, stronger would be the enhancement in the quality of localization. However, the weak longitudinal modulation favors the TL in the perturbation limit only. To support this argument, we have plotted the $\omega_{eff}$ variation for different depths of modulation when the transverse disorder strength is fixed at 0.4 in Fig. 5. In comparison with the behavior of an unmodulated disordered lattice, the counterpart $\omega_{eff}$ decreases as $\mu$ increases in a modulated disordered lattice. A stronger modulation may degrade the quality of localization. Moreover, this degradation effect may also depend on the spatial period of the modulation. Hence, the light dynamics in these modulated disordered lattices eventually synthesizes the two different categories of localization effects in the perturbation limit and may open up new possibilities to mould light propagation in such discrete photonic structures.

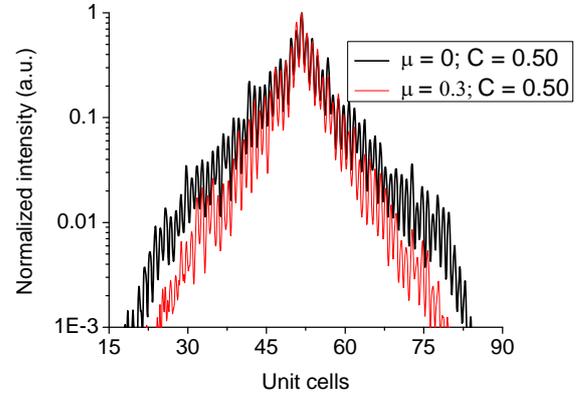

Fig. 4. (Color online) Ensemble averaged localized states: plotted in black with 50% disorder (no modulation); and in red with modulation ($\mu = 0.3$) and 50% disorder from a 15 mm long lattice when an input Gaussian beam of FWHM 10 μm was launched. The exponential nature of the decaying tails of a localized state is more prominent in the presence of modulation.

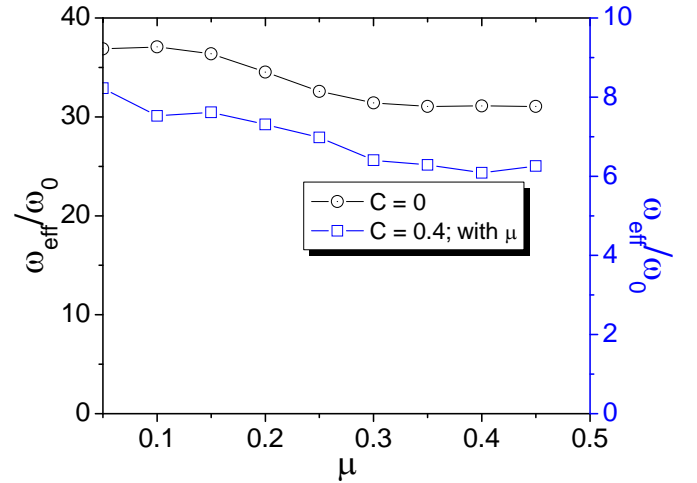

Fig. 5. (Color online) Variation in the ensemble averaged effective width of the output beam for an input Gaussian beam with different depths of modulation after a propagation distance of 15 mm: in an ordered lattice and disorder lattice with $C = 0.4$ respectively.

IV. CONCLUSIONS

Simultaneous presence of weak longitudinal modulation with transverse disorder in refractive index favors transverse localization to occur and enhances the possibility to achieve localized state of light in a shorter distance. This study would help to tune the threshold level of disorder to achieve localized state of light in disordered optical systems. The proposed platform along with the simultaneous presence of transverse disorder and dynamic modulation can be explored to realize 1D random/disordered laser.

ACKNOWLEDGMENT

This work relates to Department of the Navy Grant N62909-10-1-7141 issued by Office of Naval Research Global. The United States Government has royalty-free license throughout the world in all copyrightable material contained herein.